\documentclass{aa}  

\usepackage{graphicx}
\usepackage{txfonts}
\begin{document}

   \title{The central velocity dispersion of the Milky Way bulge
\thanks{Based on observations taken at the ESO Very Large Telescope with
the MUSE instrument under programme IDs 060.A-9342 (Science Verification; 
PI: Valenti/Zoccali/Kuijken), and 99.B-0311A (SM; PI: Valenti).}}

   \author{E. Valenti,
          \inst{1}
         M. Zoccali,
          \inst{2,3}
         A. Mucciarelli,
          \inst{4,5} 
          O. A. Gonzalez,
          \inst{6}
          F. Surot Madrid,
          \inst{1}
          D. Minniti,
          \inst{7,3,8}
	M. Rejkuba
	\inst{1,9}
	  \and
	L. Pasquini
	\inst{1}
	\and
	G. Fiorentino
	\inst{5}
	\and
	G. Bono
	\inst{10,11}
	R.~M. Rich
	\inst{12}
	M. Soto
	\inst{13,14}
          }

   \institute{European Southern Observatory, Karl Schwarzschild\--Stra\ss e 2, D\--85748 Garching 
bei M\"{u}nchen, Germany \\
              \email{evalenti@eso.org}
\and
Instituto de Astrof\'{i}sica, Pontificisa Universidad Cat\'{o}lica de Chile, Av. Vicu\~{n}a Mackenna 4860, Santiago , Chile
\and
Millennium Institute of Astrophysics, Av. Vicu\~{n}a Mackenna 4860, 782-0436 Macul, Santiago, Chile
\and
Dipartimento di Fisica e Astronomia - Universit\'{a} degli Studi di Bologna, Via Piero Gobetti 93/2, I-40129, Bologna, Italy
\and
INAF - Osservatorio di Astrofisica e Scienza dello Spazio di Bologna, via Piero Gobetti 93/3 - I-40129, Bologna, Italy
\and
UK Astronomy Technology Centre, Royal Observatory, Blacford Hill, Edinburgh EH9 3HJ, UK
\and
Departamento de Ciencias Fis\'{i}cas, Universidad Andr\'{e}s Bello, Rep\'{u}blica 220, Santiago, Chile
\and
Vatican Observatory, V00120 Vatican City State, Italy
\and
Excellence Cluster Universe, Boltzmann\--Stra\ss e 2,  D\--85748 Garching, bei M\"{u}nchen,Germany
\and
Department of Physics, University of Roma Tor Vergata, via della Ricerca Scientifica 1, 00133, Roma, Italy
\and
INAF Osservatorio Astronomico di Roma, via Frascati 33, 00040, Monte Porzio Catone RM, Italy
\and
Division of Astronomy, Department of Physics and Astronomy, UCLA, PAB 430 Portola Plaza, LA CA 90095-1547
\and
Universidad de Atacama, Departamento de F\'{i}sica, Copayapu 485, Copiap\'{o}, Chile
\and
Space Telescope Science Institute, San Martin Drive 3700, Baltimore, 21218 
             }

   \date{}

 
  \abstract
   {Recent spectroscopic and photometric surveys are providing a comprehensive view of the 
   Milky Way bulge stellar population properties with unprecedented accuracy. 
   This in turn allows us to explore the correlation between kinematics and stellar density 
   distribution, crucial to constraint the models of Galactic bulge formation.}
   {The Giraffe Inner Bulge Survey (GIBS) revealed the presence of a velocity dispersion
   peak in the central few degrees of the Galaxy  by consistently measuring high velocity dispersion in three central most fields. Due to suboptimal distribution of these fields, all being at negative latitudes and close to each other, the shape and extension of the sigma peak is poorly constrained. In this study we address this by adding new observations distributed more uniformly and in particular including fields at positive latitudes that were missing in GIBS.}
   {MUSE observations were collected in four fields at $(l, b)= (0^\circ, +2^\circ),  (0^\circ, -2^\circ)$, $(+1^\circ, -1^\circ)$, and $(-1^\circ, +2^\circ)$. 
   Individual stellar spectra were extracted for a number of stars comprised between $\sim$500 and $\sim$1200, depending on the seeing and the exposure time. 
   Velocity measurements are done by cross-correlating observed stellar spectra in the CaT region with a synthetic template, and velocity errors obtained through Monte Carlo simulations, cross-correlating synthetic spectra with a range of different metallicities and different noise characteristics.}
  {We measure the central velocity dispersion peak within a projected distance from the Galactic center of $\sim$280\,pc, reaching $\sigma V_{GC}\sim$140\,km/s at b=-1$^\circ$. 
   This is in agreement with the results obtained previously by GIBS at negative longitude.
   The central sigma peak is symmetric with respect to the Galactic plane, with
   a longitude extension at least as narrow as predicted by GIBS.

As a result of the Monte Carlo simulations we present analytical equations for the radial velocity measurement error as a function of metallicity and signal-to-noise ratio for giant and dwarf stars. 
}
   {}

   \authorrunning{Valenti et al.}  \maketitle
%
   \begin{figure*}[ht]
   \centering
\includegraphics[width=18cm,angle=0]{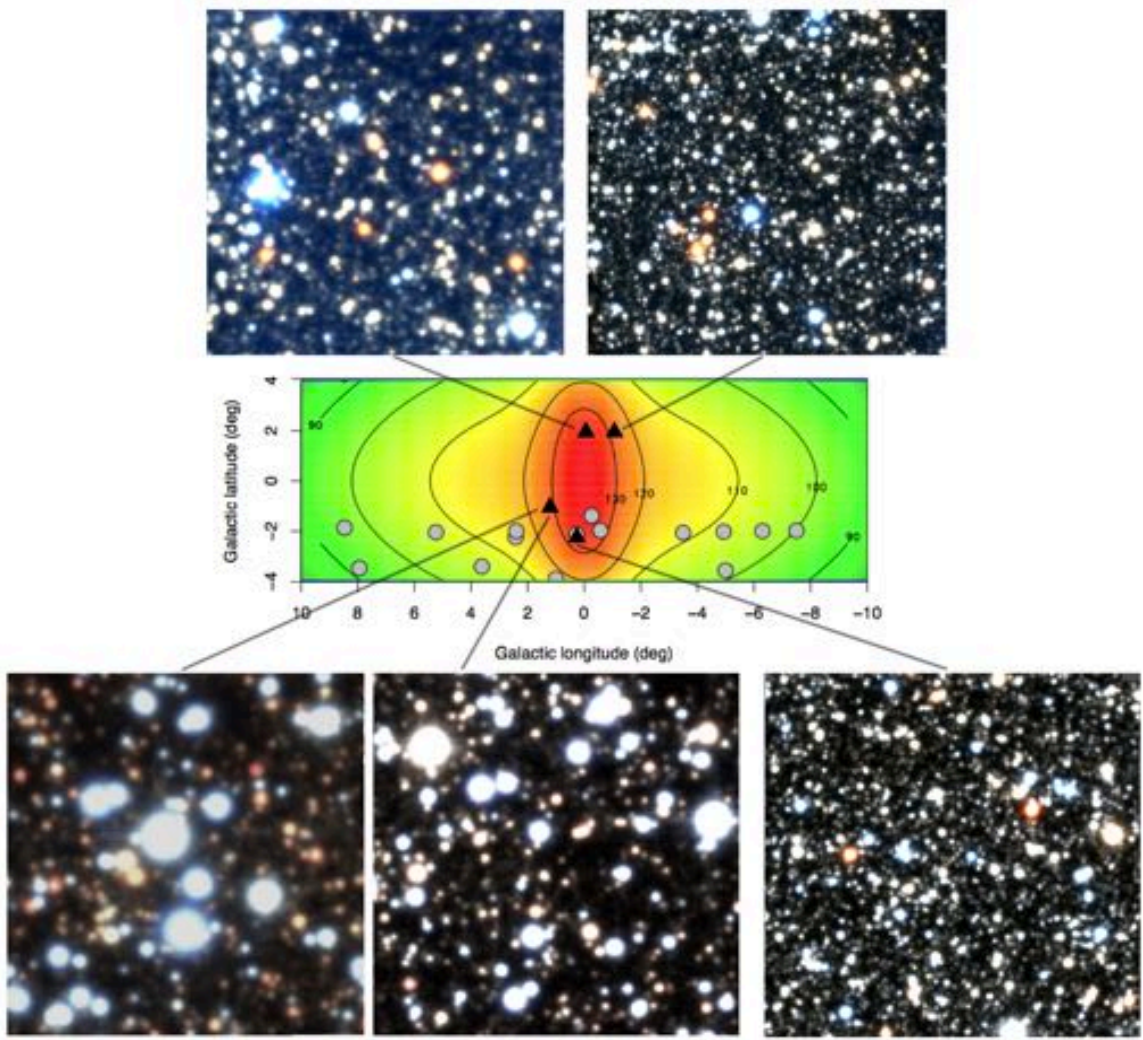}
   \caption{VRI MUSE FOV coloured images of the bulge $1'\times1'$ fields analysed here, from top left clockwise: p0p2, m1p2, p0m2, and two pointings at p1m1 highlighting the impact of different image quality. Their Galactic position relative to the velocity dispersion map of \citet{zoccali+14} is also shown.}
              \label{fig:images}
    \end{figure*}

\section{Introduction}

Recent photometric  and spectroscopic surveys of  the Galactic bulge are  providing a
wealth of data  to explore the spatial distribution, chemical  content and kinematics
of  its stellar  population.   A special  2016  edition of  the  Publications of  the
Astronomical Society  of Australia (vol.  33)  provides several reviews on  the bulgenobservational properties \citep[e.g.,][]{zoccali+16, babusiaux16}.

Stellar  kinematics  and spatial  distribution,  in  particular,  are thought  to  be
strongly correlated  with the bulge formation  process. Two main scenarios  have been
proposed for bulge formation. The first one through evolution of  the disk, when the
latter had been mostly  converted into stars.  In this case the  bulge is expected to
have  the  shape  of  a  bar,  though vertically  heated  into  a  boxy/peanut,  with
corresponding kinematics.  The second  one is  the hierarchical  merging of  gas rich
sub-clumps coming  either from the  disk or from  satellite structures. In  this case
both the spatial  distribution and the kinematics are expected  to be more isotropic.
Obviously, a combination of these two scenarios could have also led to the formation of
the Galactic bulge we observed today.

Early kinematical surveys  covering a large bulge area such  as BRAVA \citep{rich+07,
  howard+09}  and ARGOS  \citep{freeman+13,  ness+13a, ness+13b}  derived a  rotation
curve that  looked cylindrical,  supporting the  conclusion that  the bulge  had been
formed exclusively via disk  dynamical instabilities \citep{shen+10}.  These studies,
however, were  limited --  by crowding  and interstellar  extinction --  to latitudes
$|b|>4^{\circ}$.  By  using  data  from   the  GIRAFFE  Inner  Bulge  Survey  (GIBS),
\cite{zoccali+14} found that  the radial velocity dispersion ($\sigma$) exhibits a strong
increase resulting in a peak with $\sim$140 km/s,  confined within a radius of $\sim$250  pc from the Galactic
center. It was later demostrated that this  peak is spatially associated to a peak in
star counts \citep{valenti+16},  hence in stellar mass, and it  is slightly dominated
by metal poor stars \citep{zoccali+17}.

Indeed, there  is now consensus  that the inner  Galactic bulge hosts  two components
that are  best separated in  metallicity ([Fe/H]) ,  but also show  different spatial
distribution   \citep{ness+12,    dekany+13,   rojas-arriagada+14,   pietrukowicz+15,
  gran+16,zoccali+17}, kinematics \citep{babusiaux+10,  zoccali+17} and [Mg/Fe] ratio
\citep{hill+11}.

The velocity dispersion peak found from GIBS data was constrained by three fields, at
galactic coordinates  ($l,b$) = ($-0.26^{\circ}$,  $-1.40^{\circ}$) (0.27$^{\circ}$,
$-2.13^{\circ}$) and ($-0.58^{\circ}$,  $-1.98^{\circ}$), respectively. The velocity
dispersion  was derived  from samples of  441, 435  and 111  stars,
respectively.  The  need to obtain  intermediate resolution optical spectra  for many
stars, with a signal-to-noise ratio (SNR) high  enough to allow us to measure Calcium
II Triplet (CaT) metallicity, restricted the position of the GIBS innermost fields to
the  bulge hemisphere  at negative  latitudes. In  order to  constrain the  shape and
spatial extension  of the $\sigma$-peak, we  analyse here new data  obtained with the
MUSE IFU spectrograph at the ESO VLT, in  fields closer to the Galactic center at both
positive and negative latitudes.


\section{Observations and data reduction}

Three fields,  hereafter named p0m2,  p0p2 and m1p2,  located in the  innermost bulge
regions were  observed with  MUSE during the  Science Verification  campaign. Another
one, consisting of  two adjacent pointings, named p1m1-A and  p1m1-B, was observed in
Service Mode as part of a filler program 99.B-0311A (PI: Valenti) for  which only 4 hours
were executed  of the 76  hours originally approved.  Table\,\ref{tab:log}  lists the
Galactic  coordinates, exposure  times,  image quality  and interstellar  extinction
\citep{gonzalez+12} for all the fields. The two pointings of the p1m1 field were
observed under quite different seeing conditions, but they are  so close to each other that they
have the same velocity dispersion, and are thus treated as a single field hereafter.

MUSE \citep{muse} is  the integral field spectrograph  at the Nasmyth B  focus of the
Yepun (VLT\--UT4) telescope  at ESO Paranal Observatory. It provides  1 square arcmin
field of view, with a spatial pixel of 0.2", and a mean spectral resolution of $R\approx3000$.
The observations  were carried out in  seeing limited mode (WFM-noAO)  by using the
so\--called Nominal setup,  which yields a  continuous wavelength coverage between  4750$\AA$ and
9350$\AA$.

For  all  the  fields,  we  used  similar  observing  strategy  but  different  total
integration time (see Table\,\ref{tab:log}): a combination of on-target sub-exposures
each  $\sim$1000\,sec long,  taken  with a  small offsets  pattern  (i.e.  $\sim$1.5")  and
$90^\circ$  rotations in  order to  optimise the  cosmics rejection  and obtaining  a
uniform combined dataset in terms of noise properties.

The   processing  of   the   raw  data   was  performed   with   the  MUSE   pipeline
\citep[v.1.5,][]{muse-pipe}. The entire pipeline data
reduction  cascade consists  of  two  main steps:  {\it  i)}  creating all  necessary
calibrations to remove  the instrument signature from each target  exposures, such as
bias,  flats,   bad  pixels   map,  instrument  geometry,   illumination,  astrometry
correction, line spread function, response curve for flux calibration, and wavelength
solution map; and  {\it ii)} constructing, for each target  field, the final datacube
by combining the different science exposures  processed during the previous step.  In
addition  to the  final  datacube,  the pipeline  optionally  produces the  so-called
Field-of-View (FoV) images by convolving the MUSE datacube with
the transmission  curve of various filters.  For  this work we  produced FoV  images in
$V$-Johnson, $R$-Cousins  and $I$-Cousins.   Fig.\,\ref{fig:images} shows  the color
image  of   each  target  field  obtained   combining  the  $V$,  $R$   and  $I$  FoV
images, and the position of the four fields in the velocity dispersion map provided
by the GIBS survey. 
Clearly, the number  of stars detected in each field is  affected both by the
different seeing  conditions and by  the extinction of the field.

\begin{table}
\caption{Galactic coordinates, reddening, image quality and observations Log of the observed fields.
\label{tab:log} }
\centering
\begin{tabular}{lrrccc}
\hline\hline
Field & $l$~~~~ & $b$~~~~  & Exp. Time & FWHM & E($J-Ks$)\\
\hline
 p0m2    &   +0.26$^{\circ}$ & $-$2.14$^{\circ}$ & 6 $\times$ 1000\,s  &  0.6" & 0.36 \\
 m1p2    & $-$1.00$^{\circ}$ &   +2.00$^{\circ}$ & 2 $\times$ 1000\,s  &  0.5" & 0.86 \\
 p0p2    &    0.00$^{\circ}$ &   +2.00$^{\circ}$ & 3 $\times$ 1000\,s  &  1.1" & 0.90 \\
 p1m1-A  &   +1.20$^{\circ}$ & $-$1.00$^{\circ}$ & 6 $\times$ 1066\,s  &  1.2" & 0.87 \\
 p1m1-B  &   +1.20$^{\circ}$ & $-$1.00$^{\circ}$ & 6 $\times$ 1066\,s  &  0.9" & 0.86 \\
 \hline\hline
 \end{tabular}
\end{table}

\subsection{Extraction of the spectra}

The procedure adopted to  extract the spectra for all the  stellar sources present in
the target fields consists of two main steps: {\it i)} the creation of a master star
list for each field;  and {\it ii)} the reconstruction of the spectrum of each star in
the list, by using the star flux as measured in the MUSE final data cubes.

We first performed standard aperture photometry, with DAOPHOT \citep{daophot}, on the
FoV images to obtain a master list of all sources with significant counts ($>4\sigma$)
above  the background.   Due  to  the relatively  modest  crowding,  aperture or  PSF\--fitting
photometry yield virtually  identical results, therefore we  used aperture photometry
hereafter  (see  below).  Fig.\,\ref{fig:cmd}  shows  the  derived  color\--magnitude
diagrams (CMDs) for all the fields, either in  the ($R$, $V-R$) or in the ($R$, $R-I$)
instrumental plane. The latter was used for the p1m1 field because, due to its higher
extinction, the $V$ image had the lowest SNR. Here the impact of the different seeing
and  total exposure  time is  also very  clear, with  the p0p2  field being  the least
populated, due to the combination of  relatively poor seeing and short total exposure
time. The p1m1 field is the closest one to the Galactic plane, therefore showing a prominent 
disk main sequence (MS) that is both more prominent and extends to brighter magnitudes in 
comparison to the other fields.
This is  due to the fact that  at $b=-1^{\circ}$ the
optical depth of the thin disk is larger.
The presence  of bright
blue stars is very evident also in the FoV images of Fig.\,\ref{fig:images}

   \begin{figure}
   \centering
\includegraphics[width=9cm,angle=0]{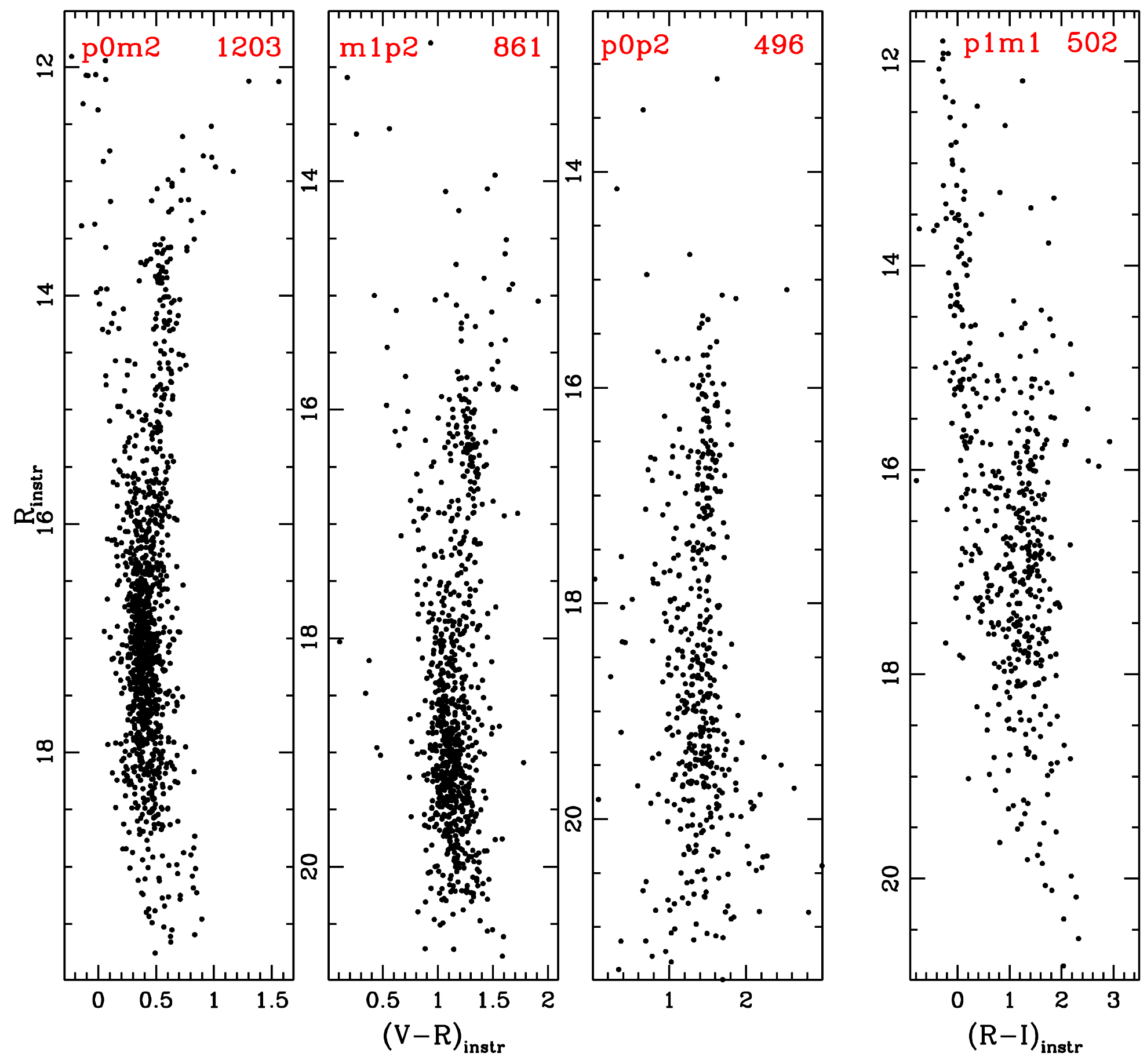}
   \caption{Instrumental CMD of the observed bulge fields as derived by running aperture photometry 
on the MUSE FoV images. The name of the fields and total number of detected stars 
are given.}
              \label{fig:cmd}
    \end{figure}

The  final MUSE  data cubes  were then  sliced along  the wavelength  axis into  3681
monochromatic images (i.e.   single planes) sampling the target  stars from 4750$\AA$
to  9350$\AA$ with  a step  in wavelength  of 1.25\,$\AA$.   Aperture photometry  was
performed on  each of these  images (task {\tt  PHOT} of {\tt  IRAF}\footnote{IRAF is
  distributed by the National Optical  Astronomy Observatories, which are operated by
  the Association of Universities for  Research in Astronomy, Inc., under cooperative
  agreement with the National Science Foundation.})  with an aperture radius $\sim$\,
1.5\,$\times \,<FWHM>$, where $<FWHM>$ is the average image quality measured over the
wavelength range $\lambda >$\,6000$\AA$. Finally, for each star in the master list,
the corresponding  spectrum was  obtained by  assigning to  each wavelength  the flux
measured on  the corresponding monochromatic image.  For a given field, a single value 
was  used for the
aperture radius, since  the FWHM variation, as measured across  the entire wavelength
range, is about half a pixel, independent from the mean FWHM value.

It is worth mentioning that this {\it  photometric} approach to the extraction of IFU
spectra has  the advantage of  successfully addressing  the issue of  sky subtraction
residuals often present  in the final data  cubes. Indeed, it is well  known that the
sky subtraction may be not always  optimal, leading to the
presence of artefacts  (e.g. weak emission line residuals  and/or {\it p\,cygni}-like
profiles) in the final spectra. By contrast, any such residuals present in the single
plane  images are  fully  taken  into account  by  the  photometric procedure,  which
estimates a local sky background for each source present in the master list.

Several attempts  at using  PSF-fitting photometry on  the monochromatic  images were
performed.  They were  finally discarded because the majority of  the stars were lost
in a few of the monochromatic images,  corresponding to the bottom of their strongest
absorption lines.  When the star flux is  close to the sky level, aperture photometry
still assigns a  meaningful flux value, while PSF-fitting photometry  just discards the
star from the list. This is not a negligible issue, given that the strong absorption
lines are very important in the measurement  of radial velocities. On the other hand,
given the modest crowding  of the images, the photometry from aperture and PSF-fitting yielded
similar quality result, at least on the FoV images. Therefore we
judged not necessary, in this case, to try and overcome the problem of non-convergence
of the PSF in the low signal regime. The analysis of more crowded fields such as the
inner regions of dense star clusters might require some different approach.

\begin{figure}
   \centering
\includegraphics[width=6.7cm,angle=270]{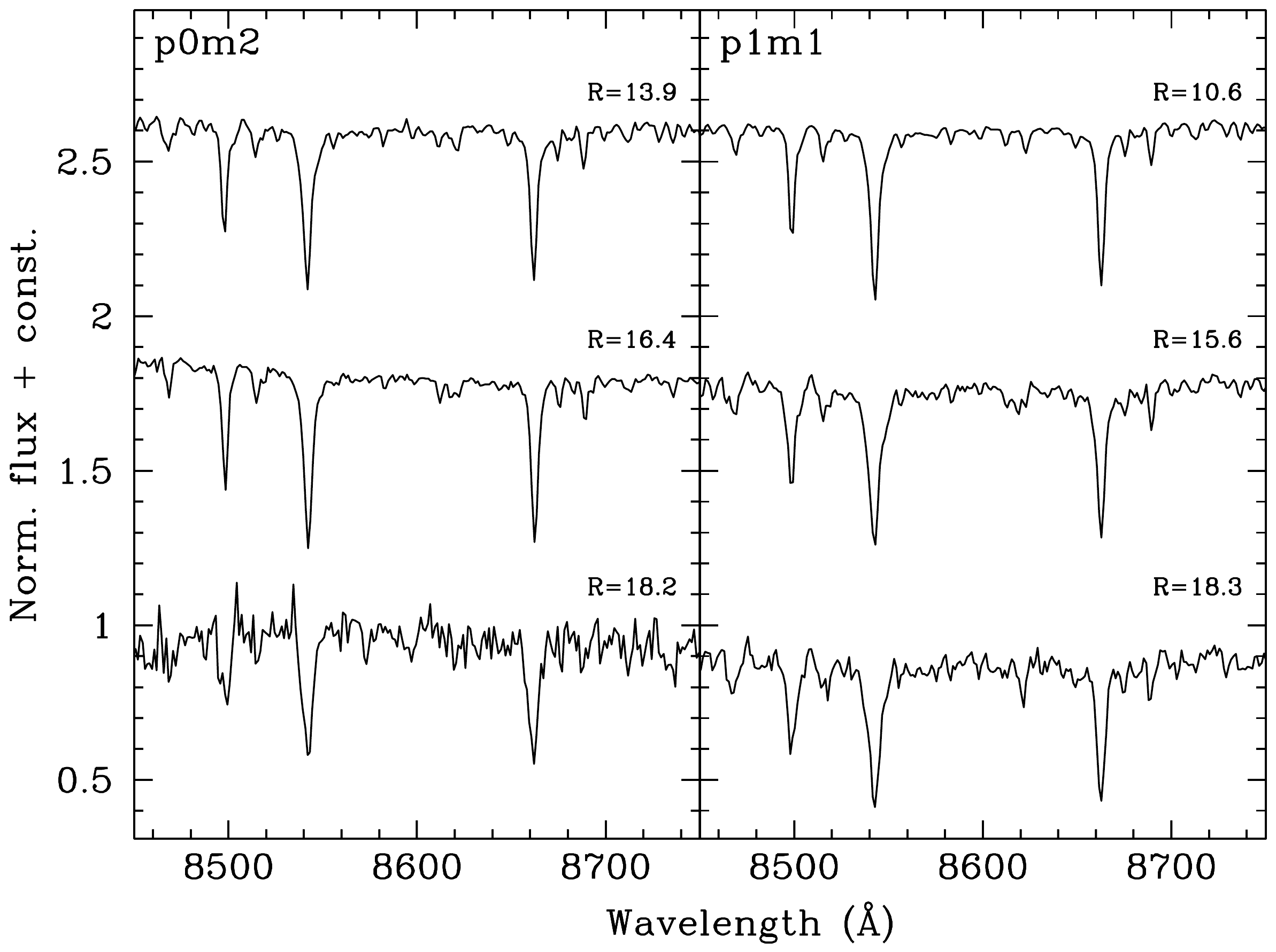}
   \caption{Example of  typical spectra in the  CaT region,  for stars in the brightest
   0.5 mag bin, for red clump stars, and star at the very faint\--end magnitude range.  
   For each spectrum, the instrumental R magnitude of the star is given.}
              \label{fig:specs}
    \end{figure}

Spectra for 1203, 861,  496 and 502 stars were reconstructed in  the p0m2, m1p2, p0p2
and p1m1 fields, respectively.  Examples of  typical extracted spectra, zoomed in the
CaT region, for  stars of different magnitudes are given  in Fig.\,\ref{fig:specs}. We
show spectra  for the  field with the  best combination of  seeing and  exposure time
(p0m2) and the  one with the worst  seeing (p1m1-A).  The  average SNR, as
measured in  the CaT  wavelength range,  of field p0m2,  m1p2, p0p2 and p1m1 stars  in the
faintest 0.5 mag bin is $\sim$\,20, 15, 10, 10 respectively.
    
 
\section{Radial velocities and velocity errors}

We  measured the  heliocentric radial velocity (RV) of  all stars  detected in  the
observed fields  through cross-correlation  with a synthetic  template by  using the
{\tt  IRAF} task  {\it  fxcor}.  Specifically,  we  adopted for  all  stars the  same
synthetic  spectra  of a  relatively  metal  rich  ([Fe/H]$=-0.4$\,dex) K  giant  and
performed the  cross-correlation between the  model and  the observed spectra  in the
wavelength range bracketing the CaT lines. To assess the  effect that  the use  of a 
single  metallicity template  may have  on the derived  velocities, in  the case  of 
p0m2  stars field,  we also  used 2  additional synthetic  templates with metallicities   
([Fe/H]$=-1.3$\,dex  and $+0.2$\,dex) that bracket the typical metallicity distribution 
function observed in the GIBS fields by \citet{zoccali+17}.
We found that the  RV  derived with the  metal\--poor and
metal\--rich models  always agree  within $\leq\,1$\,km/s, thus confirming what already
noticed by \citet{zoccali+14} that the metallicity of the adopted synthetic template
has a very minor effect on the derived RV.

 
   \begin{figure}
   \centering
\includegraphics[width=8cm]{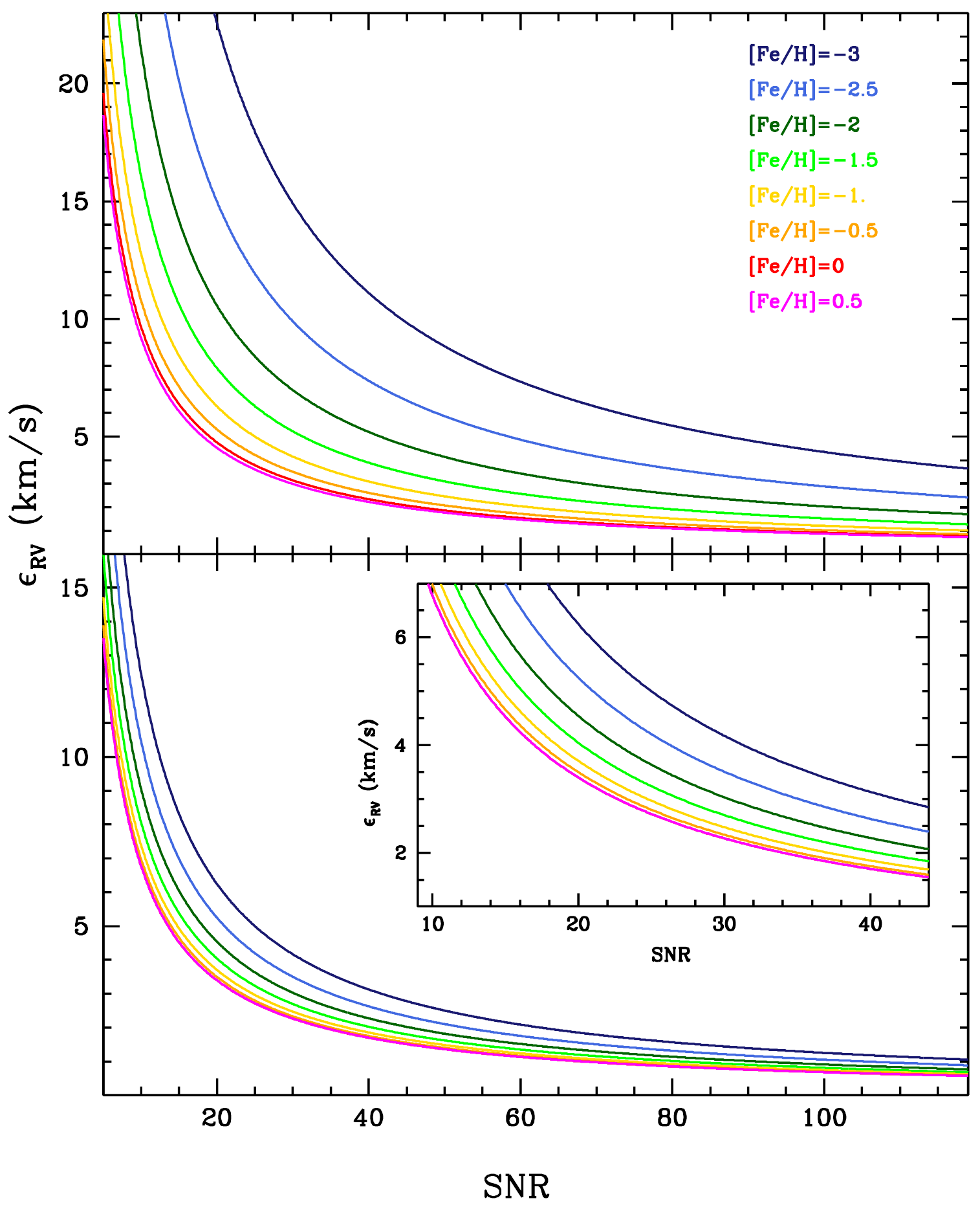}
   \caption{RV error ($\epsilon_{RV}$) profile as a function of the spectra SNR for 
   giants (bottom panel) and dwarf (top panel) of different metallicity. The inset shows
   the $\epsilon_{RV}$ profile of giants in the medium\--low SNR regime.}
              \label{fig:err_snr}
    \end{figure}
The  error in the derived RV could not be
estimated from repeated  measurements because, for each target field , we only have one
single data-cube. Therefore, the uncertainty  ($\epsilon_{RV}$) has been estimated by
means of MonteCarlo  simulations. The main sources of uncertainties  are the SNR, the
spectral resolution, and the  sampling. In order to  evaluate their impact on
the derived RVs  we generated different sets of artificial  MUSE spectra, varying SNR
and metallicity,  reproducing the observed  ones.  We started from  synthetic spectra
calculated  with   the  code  {\tt  SYNTHE}   \citep{Sbordone+04},  assuming  typical
parameters of a  giant ($T_{eff}$=~4500 K, log~g=1.5, $v_{turb}$=2 km/s)  and a dwarf
star ($T_{eff}$=~6500  K, log~g=4.5,  $v_{turb}$=1 km/s), and  considering a  grid of
metallicity  between [Fe/H]=$-$3.0  and +0.5\,dex  with  a step  of 0.5\,dex.   These
synthetic  spectra have  been  convolved with  a Gaussian  profile  to reproduce  the
spectral resolution of MUSE and then resampled at the same pixel size of the observed
extracted spectra (1.25 \AA/pixel).  Poisson noise was added to the synthetic spectra
in order  to reproduce different  noise conditions, from SNR$\sim$10  to SNR$\sim$100
with steps  of 10. At the  end, for each metallicity  and SNR, a sample of 500 synthetic
spectra  with randomly added noise was generated  and their  RVs measured  through cross-correlation  technique
({\sl fxcor}) with the original synthetic spectrum as template.

The  dispersion of  the derived  RVs of  each sample  has been  assumed as  1$\sigma$
uncertainty in the  RV measurement for a  given SNR and metallicity.   We derived the
following relations that link the radial velocity error to SNR and metallicity for giant stars (1):
\begin{equation}
$$\ln(\epsilon_{RV})=4.209 - 0.997\ln({\rm SNR})-0.029{\rm [Fe/H]}+0.058{\rm [Fe/H]}^{2}$$
\end{equation}
and for dwarf stars (2):
\begin{equation}
$$\ln(\epsilon_{RV})=4.624 - 1.023\ln({\rm SNR})-0.159{\rm [Fe/H]}+0.120{\rm [Fe/H]}^{2}$$
\end{equation}
The behaviour of $\epsilon_{RV}$ as a function  of SNR for different values of [Fe/H]
is  shown in  Fig.\,\ref{fig:err_snr} for  giant and  dwarf stars.   For giant  stars
$\epsilon_{RV}$ increases rapidly for SNR  smaller than 30, reaching uncertainties at
SNR=~10 of 6.8  and 12.5 km/s for  [Fe/H]=+0.5 and --3.0 dex,  respectively, while at
high SNR $\epsilon_{RV}$ is almost constant and  close $\sim$1 km/s.  At a given SNR,
$\epsilon_{RV}$ increases as decreasing [Fe/H], due  to the weakening  of the
CaT  lines,   while  for  [Fe/H]  larger   than  -1.0  dex  the   curves  are  almost
indistinguishable.  A  similar general behaviour is  found also for the  dwarf stars,
but with larger uncertainties because of the weakness of the CaT lines: in particular
at  SNR=~10, the  relation provides  $\epsilon_{RV}$=~9.2 km/s  for [Fe/H]=+0.5  dex,
while at lower metallicities the  uncertainties increase dramatically (up to $\sim$45
km/s for  [Fe/H]=--3.0 dex).  Note  that we limited this  procedure to the  CaT lines
spectral region, in  a window between 8450 \AA and 8700  \AA , hence these
relations are specific to the case of MUSE in this spectral window.

For each field, Table\,\ref{tab:errors} lists the errors on the RV estimates 
for stars in the faintest 0.5\,mag bin as derived by using the above mentioned relations
for two different metallicity values: [Fe/H]=--1\,dex and [Fe/H]=+0.5\,dex, which 
represent the metal\--poor and metal\--rich edge of the typical bulge metallicity
distribution function. In particular, due to the differences in the magnitude depth 
among the different fields, the values quoted in Table\,\ref{tab:errors} have been
derived by using equation (2) for the p0m2, m1p2 and p0p2 fields, whereas for p1m1 
we have adopted the relation for giants (i.e. equation (1)). 
As expected, we found that at fixed SNR (i.e. magnitude) metal\--rich 
stars have typically smaller radial velocity error. However, the variation over the entire 
bulge metallicity range is $\leq\,2$\,km/s (see Table\ref{tab:errors} and 
Fig.\ref{fig:err_snr}).
   
\begin{table}[h]
\caption{Typical radial velocity error of observed stars in the faintest 0.5\,mag bin as 
derived from equation (2) for [Fe/H]=--1\,dex (MP) and [Fe/H]=+0.5\,dex (MR).
\label{tab:errors} }
\centering
\begin{tabular}{cccc}
\hline\hline
Field & SNR & $\epsilon_{RV}$ (MP) & $\epsilon_{RV}$ (MR)\\
&&km/s&km/s\\
\hline
 p0m2    &  20 & 6.3 &  4.5 \\
 m1p2    &  15 & 8.4 & 6.1 \\
 p0p2    &  10 & 12.8 & 9.2\\
 p1m1$^a$&  10 & 7.4 & 6.8 \\
 \hline\hline
 \multicolumn{4}{l}{$^{(a)}$For this field the faintest stars are giants,}\\
 \multicolumn{4}{l}{ therefore we have used equation (1).}\\
 \end{tabular}
\end{table}

\section{Velocity dispersion}

\begin{figure}
   \centering
\includegraphics[width=7cm,angle=270]{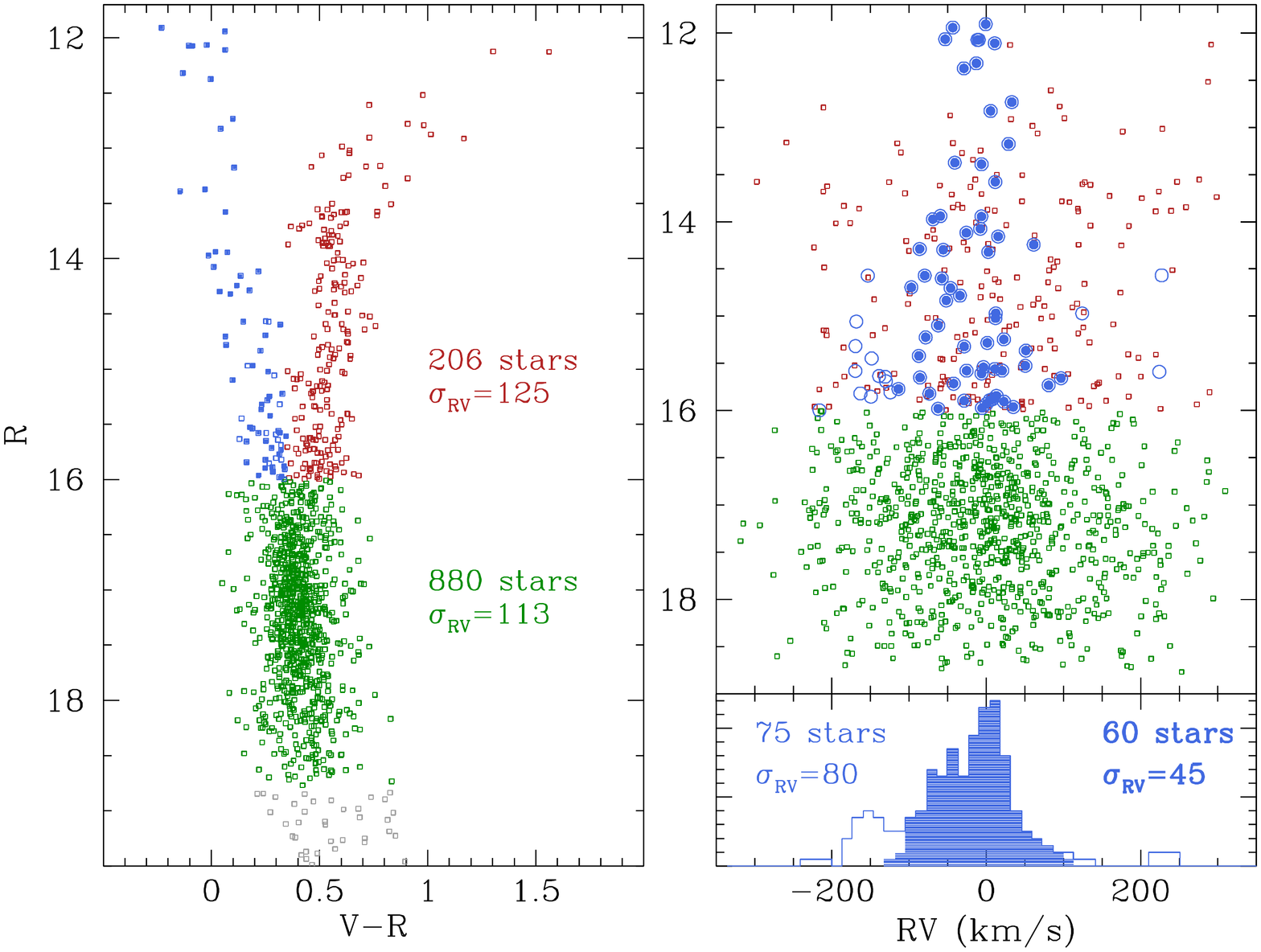}
   \caption{{\it left:} CMD  of the global targets sampled in p0m2 color\--coded according to
     their  evolutionary  phase.  Red  and  blue symbols  refer  to the  bona\--fide
     bulge\--RGB and disk\--MS  stars, respectively. For the bona\--fide sample, the
     total number of stars and their radial velocity dispersion are also given. Green
     circles mark  either bulge\-- or disk\--MS  and MS\--TO stars.
{\it  Top right:} heliocentric RV as a function  of the star
magnitude  of all stars adopting the same color  code as in the left panel. Blue  solid
symbols refer to disk\--MS stars  with $|RV|<120$ km/s.  {\it Bottom right:}
heliocentric radial velocity distribution of  disk\--MS stars. The velocity
dispersion  of the total disk\--MS  sample and  of  the sub\--sample  obtained after
applying  a cut  at $|RV|<120$ km/s are given.}
              \label{fig:p0m2Sigma}
    \end{figure}

\begin{figure}
   \centering
\includegraphics[width=7cm,angle=270]{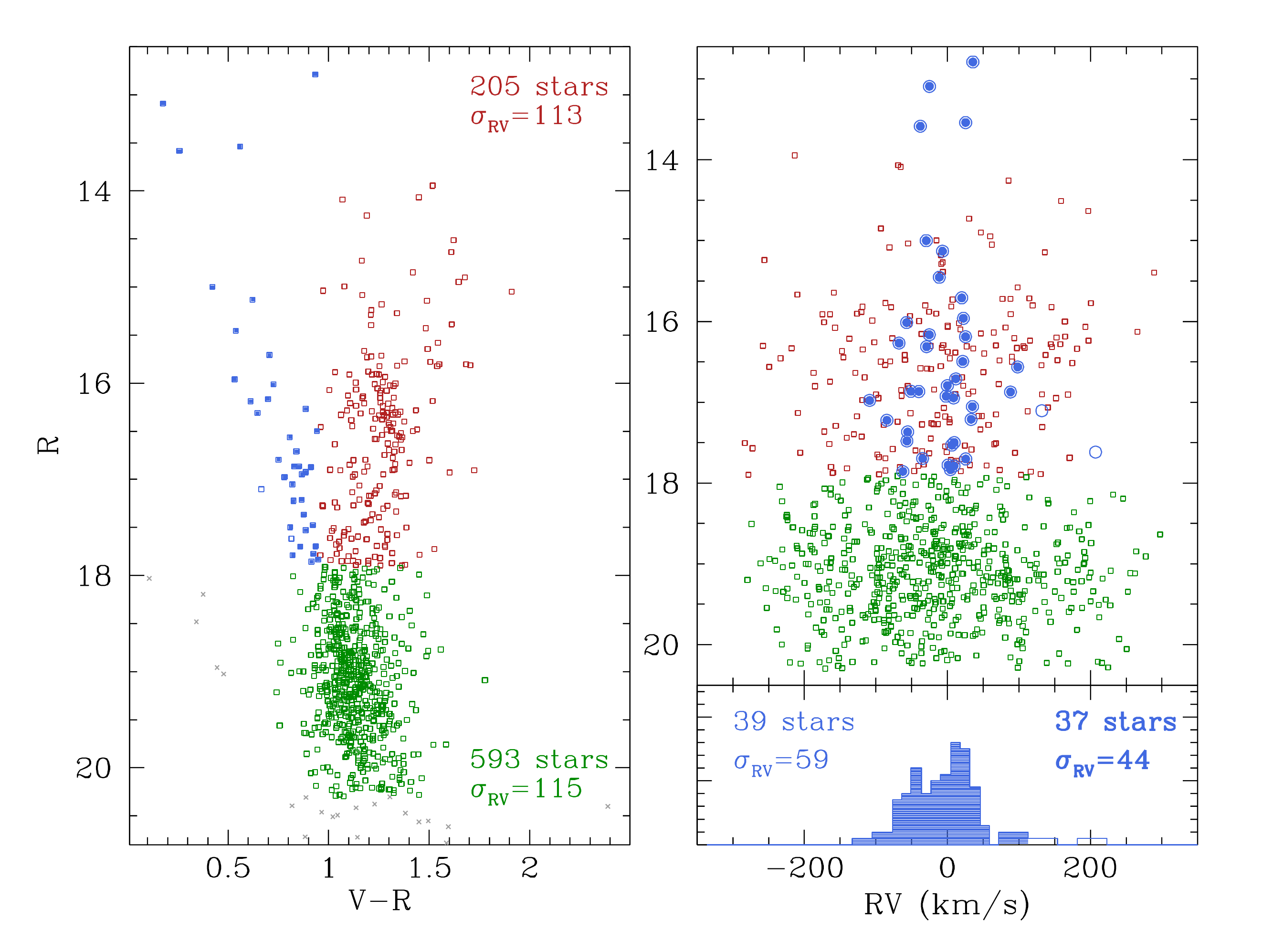}
   \caption{Same as Fig.\,\ref{fig:p0m2Sigma} for m1p2 field.}
              \label{fig:m1p2Sigma}
    \end{figure}

\begin{figure}
   \centering
\includegraphics[width=7cm,angle=270]{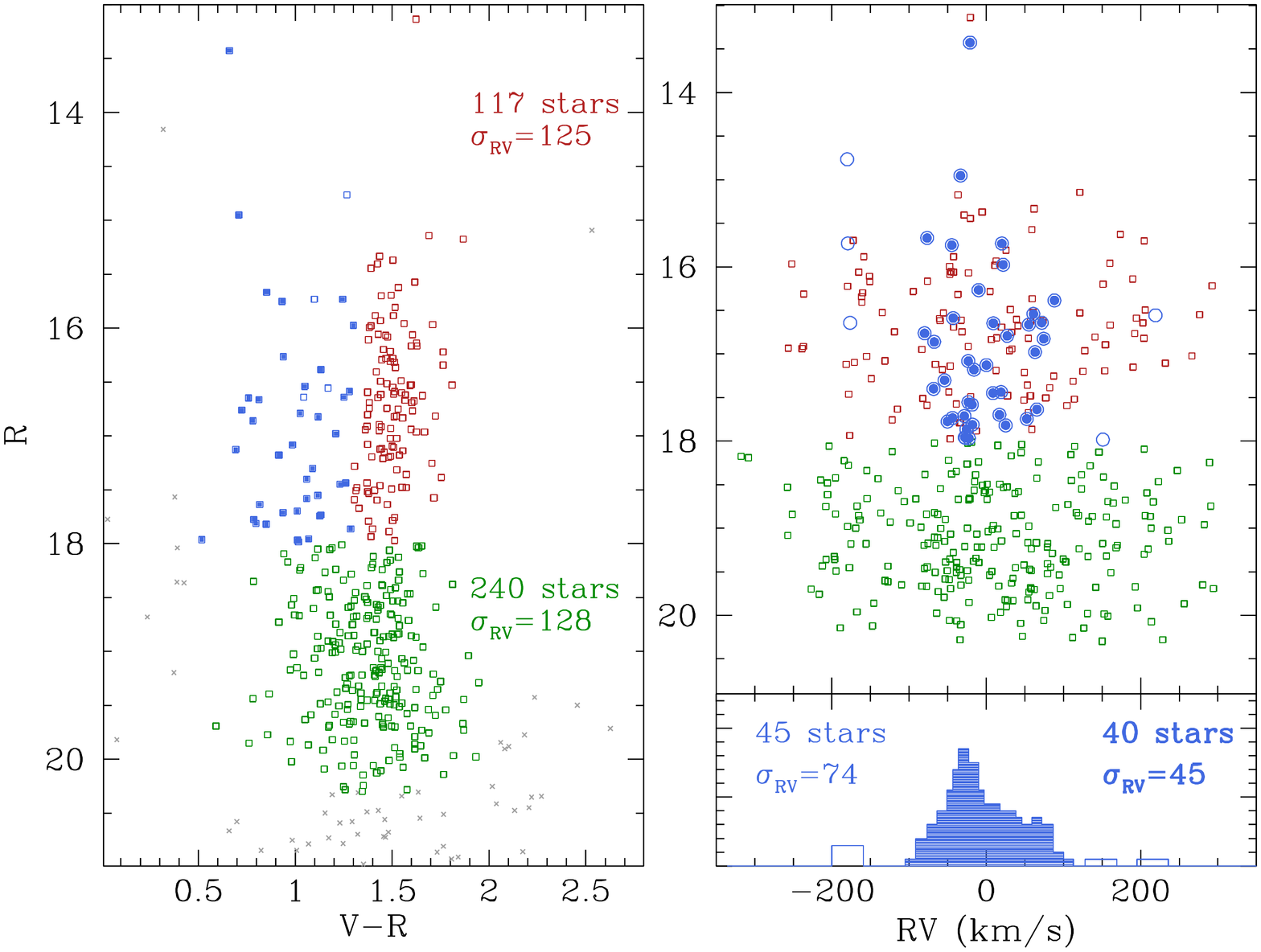}
   \caption{Same as Fig.\,\ref{fig:p0m2Sigma} for p0p2 field.}
              \label{fig:p0p2Sigma}
    \end{figure}

\begin{figure}
   \centering

\includegraphics[width=6.7cm,angle=270]{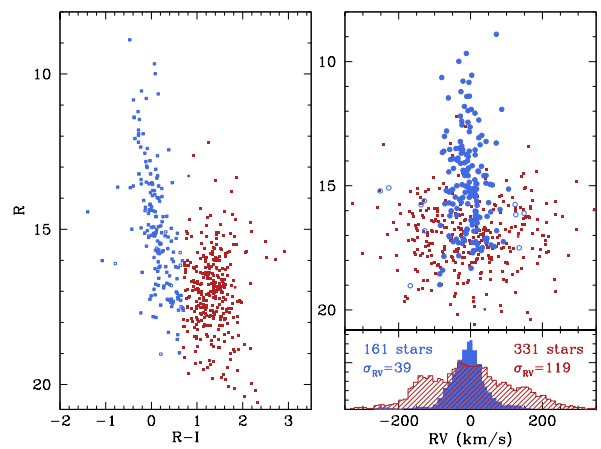}
   \caption{Same as Fig.\,\ref{fig:p0m2Sigma} for p1m1 field. In this
case the bulge MS
            falls below the limit magnitude, and therefore bulge RGB
stars can be separated
            from disk MS stars by means of a simple color cut.}
              \label{fig:p1m1Sigma}
    \end{figure}

In order to measure  the velocity dispersion of bulge stars, it  is important to take
into account  the contamination  by foreground  disk stars, which are known to have  a smaller
velocity  dispersion  \citep{Ness+16,Robin+17}.   
An estimate  of the  actual  disk velocity  dispersion 
can  be attempted by  selecting foreground   disk  MS   stars  in the  instrumental CMD of each observed field. 
This   is 
shown   in
Figs.~\ref{fig:p0m2Sigma}-\ref{fig:p1m1Sigma}.   Thanks to  the good
seeing,  longer
exposure time and relatively low reddening, the  p0m2 field has the best
defined CMD,
reaching  fainter magnitudes  (Fig.~\ref{fig:p0m2Sigma}).  In  this
field,  bona fide
bulge-RGB/disk-MS stars (red/blue symbols, respectively)  are selected
as having both
$R<16$ and  $(V-R)$ larger/smaller than  $0.35$, respectively. The cuts 
isolate 75
disk MS stars, shown in blue in the CMD, and 206 bulge RGB stars shown
in red.  Stars
fainter  than  $R=16$,  plotted  in  green,  cannot  be  safely
assigned  to  either
population.   The top-right  panels of
Figs.~\ref{fig:p0m2Sigma}-\ref{fig:p1m1Sigma}
shows the heliocentric radial velocity versus  magnitude, for all the
stars, with the
same color coding  as before.  It is clear  that disk MS stars have a
lower velocity
dispersion, but, as  expected, their radial velocity distribution is
contaminated by
bulge stars, both  blue stragglers and sub  giant branch stars.  In
fact, the radial
velocity histogram shown at the bottom of  the right panel clearly shows
the presence
of outliers at $|RV|>150$ km/s. Indeed, if  these stars are excluded, by
a simple cut
at $|RV|<120$ km/s,  the radial velocity dispersion drops to  a value of
$\sigma_{\rm
  RV}=45$ km/s, consistent in all three  fields at $b=\pm2^{\circ}$.

This exercise allows us  to conclude that in the region of the  CMDs above the old MS
turnoff, where we can safely separate disk  foreground from bulge stars by means of a
color cut, the  velocity dispersion of the  disk is significantly lower  than that of
the bulge. Therefore, in order to include bulge  MS stars in our analysis, we need to
allow for the presence of two components with different kinematics.

In the field at $b=-1^{\circ}$ the data do  not reach the bulge MS, and therefore the
foreground  disk MS  and the  bulge RGB  can be  separated by  just a  color cut,  at
$(R-I)_{\rm inst}=0.7$.  In Fig.\,\ref{fig:p1m1Sigma}, the  panels on the  right show
that disk  stars, with the  same selection imposed  for the other  fields ($|RV|<120$
km/s), have a  velocity dispersion of 39 km/s.  This value is lower than  the 45 km/s
found  at   $b=\pm2^{\circ}$,  consistent   with  the  fact   that  this   field,  at
$b=-1^{\circ}$,  samples  more  thin  disk stars,  having  a  smaller  velocity
dispersion.   Bulge  RGB  stars,  on  the other  hand,  have  a  velocity  dispersion
$\sigma=119$  km/s.  This  value  will not  be further  refined,  because it  already
includes all the bulge stars measured  in this field, with a negligible contamination
from disk stars.

\begin{figure}
   \centering
\includegraphics[width=8.5cm,angle=0]{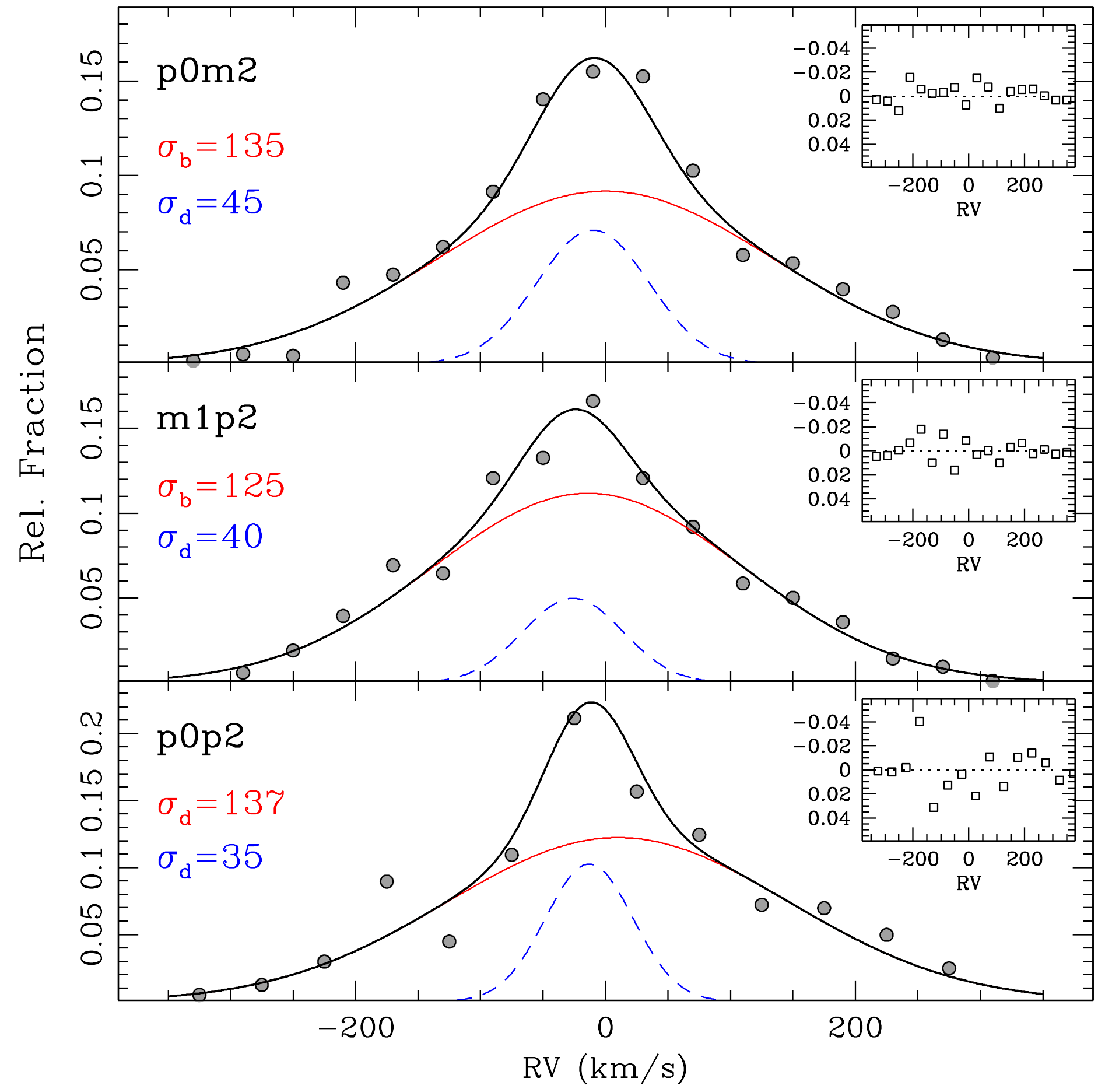}
   \caption{Normalised  heliocentric radial  velocity distribution
function for  all sampled stars  observed in the  p0m2 (top panel),  m1p2 (middle
panel)  and p0p2 (bottom panel) fields. The best\--fit  to the velocity
distribution (solid black line) is  obtained by using  a combination of two  gaussian (red
and  blue solid lines) functions whose  sigma are also reported in each  panel.
The residuals of the best\--fit are shown for each field in the insets.}
              \label{fig:Sigma}
    \end{figure}

The RV distributions  for all  the sampled  stars, in the 3  fields at
$b=\pm2^{\circ}$, are shown in Fig.\,\ref{fig:Sigma}.   As demonstrated above, 
the  best-fit to  the  observed  velocity  distribution in all
fields is obtained  with  a
combination  of  two  Gaussian  components with approximately the same  mean
($(RV_b-RV_d)\sim$10\,km/s) but  very different  $\sigma$. The foreground  disk stars
contaminating the  bulge sample have a velocity  distribution with smaller dispersion
(blue dashed line in Fig.\,\ref{fig:Sigma}).  Specifically, we found the velocity
dispersion of the disk population along the line of sight towards p0m2, m1p2 and p0p2
fields to  be $\sigma$=45, 40,  35\,km/s, respectively.  These values agree  with the
APOGEE findings  \citep{Ness+16} for disk stars  in the foreground of the bulge, within
a distance of 3 kpc,  as well  as those  reported by \citet{Robin+17} for  thick disk
stars  in the  solar  neighbourhood.  In Table\,\ref{tab:results} we list the  mean
heliocentric RV and velocity dispersion  measured for the bulge stars as
obtained from  the best-fit to the  velocity distribution of the global sample. In
addition,  following the  prescription  of \citet{ness+13b},  we provide the  mean
galactocentric radial velocity ($V_{GC}$) by correcting the  mean heliocentric
value for the Sun motion with respect to the Galactic center.

\begin{table}
\caption{Mean heliocentric and galactocentric radial velocity, and
velocity dispersion
measured for the bulge stars in the observed fields\label{tab:results} }
\centering
\begin{tabular}{cccccr}
\hline\hline
Field & $<V_{Helio}>$ & $<V_{GC}>$ & $\sigma$  &N$_{b}$/N$_{tot}$ N$_{tot}$\\
& (km/s)& (km/s)& (km/s)& (\%) &\\
 \hline
 p0m2  &   0 $\pm$ 4.4  &  +9.8  & 135 $\pm$ 3.1 & 79.5 & 1203\\
  m1p2  & $-$14 $\pm$ 4.6  &  $-$8.8  & 125 $\pm$ 3.3  & 87.5& 861\\
 p0p2  &  10 $\pm$ 7.5  &  19.2  & 137 $\pm$ 5.3 & 82.4&  496\\
 p1m1  &   1 $\pm$ 6.6  &  14.7  & 119 $\pm$ 4.7 & 67.3& 502\\
 \hline\hline
 \end{tabular}
\end{table}

   \begin{figure}
   \centering
\includegraphics[width=9cm,angle=0]{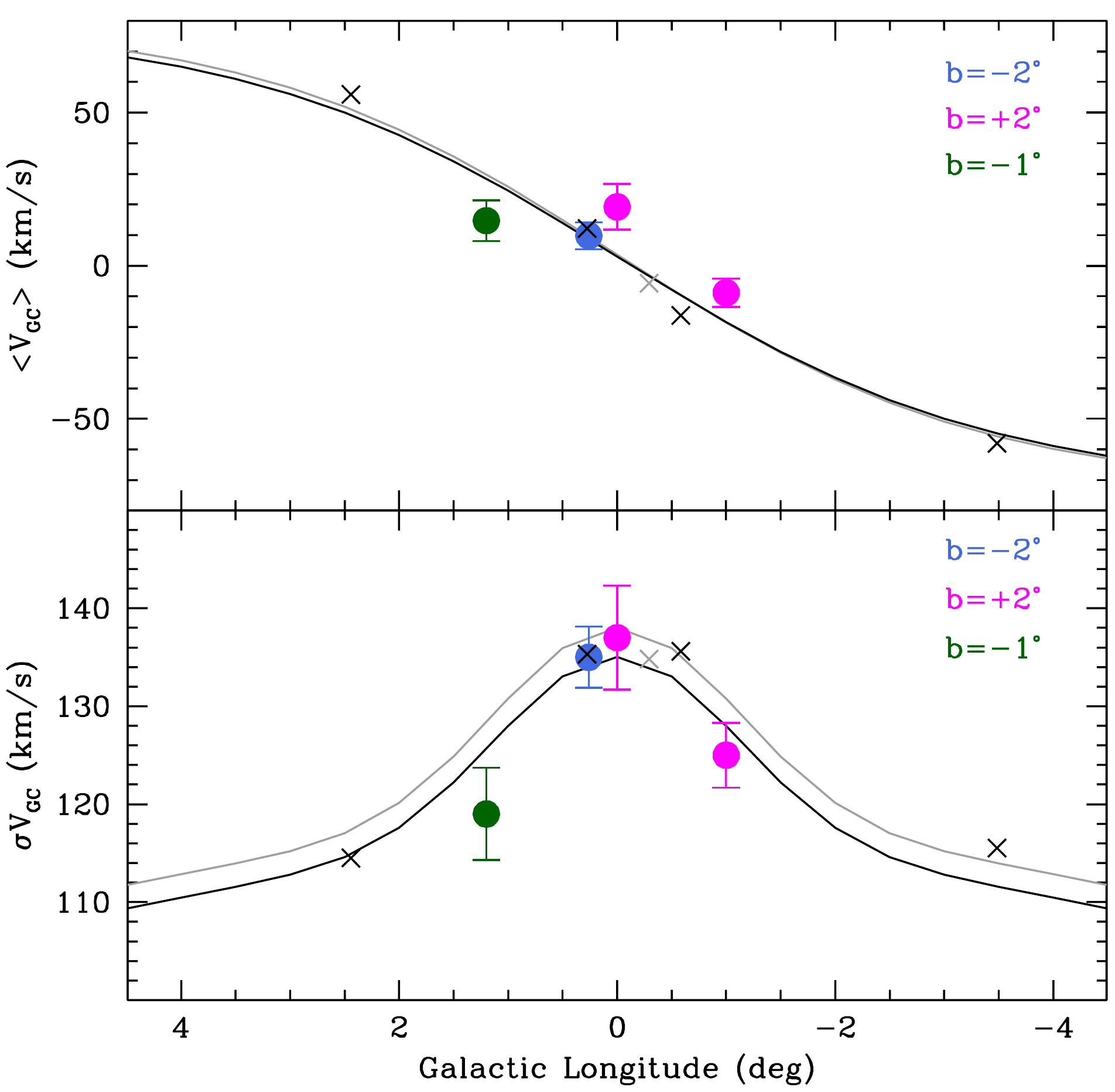}
   \caption{Mean galactocentric  radial velocity ({\it top})  and
velocity dispersion ({\it bottom}) as a function of the Galactic longitude,  for
different latitude
     as listed in the  labels. Big color symbols refer to  fields
observed with MUSE,
     whereas small crosses mark the innermost  GIBS fields at
$b=-2^{\circ}$ (black) and
     $b=-1^{\circ}$  (grey).  The  solid lines  represent the expected
trend  of the
     radial    velocity   and    velocity   dispersion according to the
maps derived
     in \citet[][equations 1 and 2]{zoccali+14}. Error bars are derived from the sampling.}
              \label{fig:SigmaConfr}
    \end{figure}

\begin{figure}[h]
\centering
\includegraphics[width=8cm]{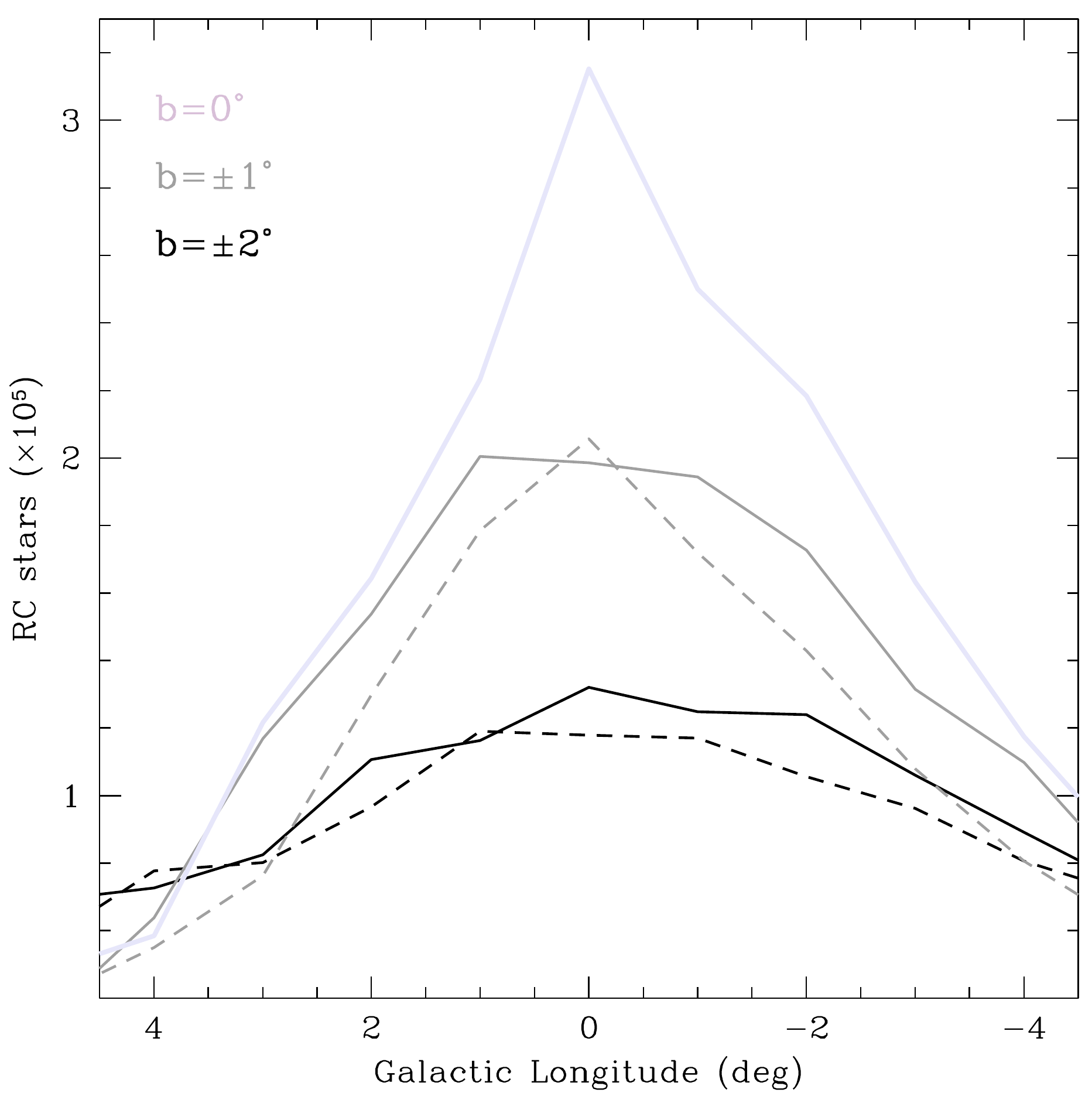}
   \caption{Profile of the  stellar projected density, as traced by RC stars, in the
     inner few degrees of the Galactic bulge, from \citet{valenti+16}. The figure is a
     zoom of their Fig.~4, with solid/dashed lines referring to negative/positive
     latitudes, respectively.}
              \label{fig:Nstars}
\end{figure}

\section{Discussion and Conclusions}

We have measured radial  velocities for several hundreds bulge stars  in each of four
fields  located   within  $l=\pm1.5^{\circ}$  and $b=\pm2^{\circ}$,  with  the  IFU
spectrograph MUSE@VLT.  All the fields are  confined within a projected radius of 280 parsecs
from the Galactic center, assuming the latter at 8 kpc from the Sun.

The aim of  this work is to assess  the presence of a large
peak in velocity dispersion  in this inner region of the  Galaxy, 
previously identified   by   \citet{zoccali+14} based on GIBS
survey, and to constrain its shape (see Fig.~\ref{fig:images} for a  zoom of the inner region of  the velocity
dispersion map derived  in  that  work).   Figure~\ref{fig:SigmaConfr} (bottom) shows 
the  central velocity dispersion  peak as measured  in the five  innermost GIBS
fields  (black and gray small points, at $b=-2^{\circ}$ and  $b=-1^{\circ}$, respectively) 
and as predicted in other Galactic positions according to the interpolated surface derived
in that paper (black and grey curves). 
The curve was  assumed to be symmetric above and below the Galactic  plane,  therefore the  
prediction  for  negative  or positive latitude  is identical by definition.  
The upper panel of Fig.~\ref{fig:SigmaConfr} shows the same but for the radial velocity.

The new values derived in the present work are plotted in Fig.\,\ref{fig:SigmaConfr} with large colored symbols. They
confirm both  the presence of  the central velocity  dispersion peak, and its absolute
value, reaching $\sigma V_{\rm  GC}$ $\sim$ 140 km/s at its center.  
We also confirm that the  peak is symmetric  above and  below the plane,  as the two 
measurements at ($l,b$)=(0$^{\circ}$,-2$^{\circ}$) and  (0$^{\circ}$,+2$^{\circ}$) are  
mutually consistent.
With  the present data we cannot constrain the latitude extension of the peak better than
what was done in  \cite{zoccali+14}, who found  that  the peak  would disappear at  the
latitude of Baade's  window ($b=-4^{\circ}$).  
We can however constrain the longitude extension of the peak,  which we show to be at 
least as  narrow as predicted by GIBS in longitude. 
In fact, the new  fields at $l=\pm1^{\circ}$ have a velocity dispersion
that is lower than the prediction of the GIBS maps.

In \citet{valenti+16} we have  derived maps of  stellar projected density and  stellar mass
from star counts  in the VVV PSF catalogues,  using red clump stars as tracers 
of the total number of  stars (and stellar  mass). We  found the presence  of a
peak  in stellar density  in  the  inner few  degrees  of  the  Galaxy,  that is
reproduced  here  in Fig.~\ref{fig:Nstars}.  This  demonstrates  that   the  sudden increase 
in  velocity dispersion is likely due to the presence of a large concentration of
stars (/mass) in the inner  Galaxy. The Galactic  position and spatial  extension of the 
peak roughly coincides  with the  sigma peak  characterised here,  even if  its
detailed  shape is somehow  different. In  particular, while  the sigma  peak is  still
rather  sharp at $b=\pm2^{\circ}$,  the  peak  in  star  counts  is  already  very
shallow  at  these latitudes. 
On  the other hand, while  the conversion from observed velocity and mass
requires dynamical modelling, as it requires  hypothesis on the orbit distribution and
their possible  anisotropy, it is  qualitatively expected that  the effect of  a mass
concentration on  the stellar velocity  is felt down to  some distance from  the mass
source. Hence  it is not  surprising that the sigma  peak is more spatially extended
than the stellar density peak.

One   thing  that   deserves  further   study  is   the  fact that, according   to
\citet{zoccali+17},  metal\--poor  stars  slightly dominate  the stellar density  at
$b=1^{\circ}$ (their  Fig.~7), but the velocity  dispersion is higher for  metal\--rich
stars, in the same field (their Fig.~12).  This is a clear evidence that the conversion
between  velocity  dispersion and  mass  involves  at  least another 
parameter  (the anisotropy of the orbit distribution) and  that this parameter is
different for metal\--poor and metal\--rich stars.

The data provided here should be included in the (chemo)-dynamical
models of the Galaxy \citep[e.g.,][]{dimatteo+15, debattista+17, portail+17, fragkoudi+18} in order to properly
take into account the mass distribution of the inner few degrees of the Milky Way.

\begin{acknowledgements}
EV, MZ, OAG, DM and MR gratefully acknowledge the Aspen Center for Physics where this
work  was partially  completed. The  Aspen  Center for  Physics is  supported by  the
National Science Foundation  grant PHY\--1066293. During their stay in  Aspen, MZ and
DM were partially  supported by a grant  from the Simons Foundation.   
Support for MZ and  DM  is  provided by  the  BASAL  CATA  Center  for Astrophysics  and  
Associated Technologies through grant PFB-06, and the Ministry for the Economy, 
Development, and Tourism's Programa Iniciativa  Cient\'\i fica Milenio through grant  IC120009, 
awarded to Millenium Institute of Astrophysics (MAS). 
EV and MZ also acknowledge support from FONDECYT Regular 1150345.
DM acknowledges support from FONDECYT Regular 117012.
\end{acknowledgements}

\bibliographystyle{aa}
\bibliography{mybiblio}
\end{document}